\documentclass[fleqn,usenatbib,useAMS]{mnras}

\usepackage{newtxtext,newtxmath}

\usepackage[T1]{fontenc}
\usepackage{ae,aecompl}

\usepackage[dvipdfmx]{graphicx}	
\usepackage{amsmath}	
\usepackage{amssymb}	
\usepackage{bm}		
\usepackage{pdflscape}	
\usepackage{color}
\usepackage{hyperref} 

\title[Constrains on ultra-low-frequency GWs] {Constraints on ultra-low-frequency gravitational waves with statistics of pulsar spin-down rates}

\author[H., Kumamoto. et al.]{
Hiroki Kumamoto$^{1, 2}$\thanks{E-mail: 171d9003@st.kumamoto-u.ac.jp},
Yuya Imasato$^{1}$,
Naoyuki Yonemaru$^{1, 2}$,
Sachiko Kuroyanagi$^{3}$,
\newauthor
and Keitaro Takahashi$^{1,4}$.\\
$^{1}$Kumamoto University, Graduate School of Science and Technology, Japan \\
$^{2}$CSIRO Astronomy and Space Science, PO Box 76, Epping NSW 1710, Australia\\
$^{3}$Nagoya University, Graduate School of Science, Japan\\
$^{4}$International Research Organization for Advanced Science and Technology, Japan
}

\date{Accepted 2019 August 18. Received 2019 August 18; in original form 2019 March 4}

\pubyear{2019}

\begin{document}
\label{firstpage}
\pagerange{\pageref{firstpage}--\pageref{lastpage}}
\maketitle
\begin{abstract}
We probe ultra-low-frequency gravitational waves (GWs) with statistics of spin-down rates of milli-second pulsars (thereafter MSPs) by a method proposed in our prevous work (Yonemaru et al. 2016). The considered frequency range is $10^{-12}{\rm Hz} \lesssim f_{\rm GW} \lesssim 10^{-10}$Hz
. 
The effect of such low-frequency GWs appears as a bias to spin-down rates which has a quadrupole pattern in the sky. We use the skewness of the spin-down rate distribution and the number of MSPs with negative spin-down rates to search for the bias induced by GWs. Applying this method to 149 MSPs selected from the ATNF pulsar catalog, we derive upper bounds on the time derivative of the GW amplitudes of $\dot{h} < 6.2 \times 10^{-18}~{\rm sec}^{-1}$ and $\dot{h} < 8.1 \times 10^{-18}~{\rm sec}^{-1}$ in the directions of the Galactic Center and M87, respectively. Approximating the GW amplitude as $\dot{h} \sim 2 \pi f_{\rm GW} h$, the bounds translate into $h < 3 \times 10^{-8}$ and $h < 4 \times 10^{-8}$, respectively, for $f_{\rm GW} = 1/(1000~{\rm yr})$. Finally, we give the implications to possible super-massive black hole binaries at these sites.
\end{abstract}

\begin{keywords}
gravitational waves -- methods: data analysis -- methods: statistical -- pulsars: general.
\end{keywords}



\section{Introduction}
\label{section1}

Laser Interferometer Gravitational Wave Observatory (LIGO) has succeeded in detecting gravitational waves (GWs) with frequencies $\sim 100~{\rm Hz}$ radiated from black hole binaries with masses of about $30 M_{\odot}$ \citep{LIGO}. With KAGRA \citep{KAGRA} and VIRGO \citep{VIRGO}, ground-based interferometers will pave the way for gravitational-wave astronomy. In fact, this is the first step toward multi-wavelength gravitational-wave astronomy and lower-frequency GWs are to be probed: space interferometers such as Laser Interferometer Space Antenna (LISA) \citep{LISA} and DECIGO \citep{DECIGO}, Pulsar Timing Arrays (PTAs) such as Parkes Pulsar Timing Array (PPTA) \citep{PPTA}, European Pulsar Timing Array (EPTA) \citep{EPTA}, North American Nanohertz Observatory for Gravitational Waves (NANOGrav) \citep{NANOGrav} and International Pulsar Timing Array (IPTA) \citep{IPTA}, and observations of B-mode polarization of the cosmic microwave background such as Simons Observatory \citep{Simons} and LiteBIRD \citep{LiteBIRD}.

GWs radiated from possible super-massive black hole (SMBH) binaries in the Galactic Center (GC) will greatly improve our understanding of gravity theories, astrophysics of SMBHs and environment of the GC region. In the late stage of SMBH binary evolution, GWs with frequencies of $10^{-9}~{\rm Hz} \lesssim f_{\rm GW} \lesssim 10^{-6}~{\rm Hz}$ are radiated and the range is main target of PTAs. The frequency range of PTA is determined by the observational time span and cadence, and lower frequencies $f_{\rm GW} \lesssim 10^{-10}~{\rm Hz}$ are difficult to probe.
In fact, the sensitivety is expected to scale as $f_{\rm GW}^{-2}$ toward lower frequencies \citep*{Blandford, Moore}.

In \cite{Yonemaru1}, we proposed a new detection method for GWs with ultra-low-frequencies of $f_{\rm GW} \lesssim 10^{-10}~{\rm Hz}$ (for other methods, see \cite{Bertotti,Kopeikin1,Potapov}). The method utilizes the fact that the spin-down rate of milli-second pulsars (MSPs) is biased by such GWs, since both give the same quadratic time dependence to the time of arrival of pulses. This effect depends on the relative direction of the GW source and a pulsar. Thus, statistics of the spin-down rate distributions can probe such GWs as we describe later. In \citet{Yonemaru2}, by using simulated 3,000 MSPs, which are expected to be discovered by the Square Kilometre Array (SKA) survey, we estimated the sensitivity of this method in a simple situation, where we assume that MSPs are located uniformly in the sky and the ``pulsar term'' is neglected. We concluded that GWs with the derivative of amplitude as small as $3 \times 10^{-19}$ s$^{-1}$ could be detected. Then, in \cite{Hisano}, we considered a more realistic model of MSP distribution in the Galaxy and took the pulsar term into account in order to obtain more accurate estimates of the sensitivity by extending the analysis of \cite{Yonemaru2}. We found that the sensitivity depends on the direction, polarization and frequency of GWs and becomes worse at low frequencies ($f_{\rm GW} \lesssim 10^{-12}~{\rm Hz}$) because of the pulsar term.

This work is the first attempt to apply the above method to real data. We use MSPs selected from the current ATNF pulsar catalog \citep{PSRCAT} and probe GWs with a frequency of $10^{-12}~{\rm Hz} \lesssim f_{\rm GW} \lesssim 10^{-10}~{\rm Hz}$.
In Section \ref{section2}, we give a brief summary on our method proposed in \citet{Yonemaru1}. Then, after describing our data set, upper bounds on the derivative of GW amplitude are derived in Section \ref{section3}. In Section \ref{section4}, we discuss the implication of the upper bounds to possible SMBH binaries at the Galactic Center and M87. Finally, our results are summarized in Section \ref{section5}.

\section{Detection Principle}
\label{section2}

Let us first describe the detection method of ultra-low-frequency GWs following \citet{Yonemaru1}. Timing residuals induced by GW are given by \citep{Detweiler},
\begin{eqnarray}
r_{GW}(t) &=& \sum_{A = + , \times} F^A(\hat{\Omega}, \hat{p}) \int^t \Delta h_A (t^{\prime}, \hat{\Omega}, \theta) dt^{\prime},
\label{e1}
\end{eqnarray}
where we denote the direction of pulsar as $\hat{p}$, the propagation direction of GW as $\hat{\Omega}$ and the GW polarization angle as $\theta$. 
Here, antenna beam pattern $F^A(\hat{\Omega}, \hat{p},\theta)$ is the geometric factor written by \citep{Anholm},
\begin{eqnarray}
F^{A}(\hat{\Omega},\hat{p}) &=& \frac{1}{2}\frac{\hat{p^i}\hat{p^j}}{1+{\hat{\Omega}} \cdot {\hat{p}}}e^{A}_{ij}(\hat{\Omega}),
\label{e2}
\end{eqnarray}
where $e^{A}_{ij}(\hat{\Omega})$ $(A = + ,\times)$ are the GW polarization tensor given by 
\begin{eqnarray}
e^{+}_{ij}(\hat{\Omega}) &=& {\hat{m_i}}{\hat{m_j}}-{\hat{n_i}}{\hat{n_j}}, \\
e^{\times}_{ij}(\hat{\Omega}) &=& {\hat{m_i}}{\hat{n_j}} + {\hat{n_i}}{\hat{m_j}},
\label{e4}
\end{eqnarray}
with $\hat{m}$ and $\hat{n}$ being the polarization basis vectors.
In Eq.(\ref{e1}), $\Delta h_A (t^{\prime}, \hat{\Omega},\theta)$ is the difference of geometric perturbation between the earth and the pulsar. This is given by,
\begin{eqnarray}
\Delta h_A (t^{\prime}, \hat{\Omega}, {\theta}) &=& h_A (t, \hat{\Omega}, {\theta}) - h_A (t_p, \hat{\Omega}, \theta),
\label{e11}
\end{eqnarray}
where $t_p = t - \tau$ with $\tau =  L/c(1+\hat{\Omega} \cdot \hat{p})$ being the pulse propagation time from the pulsar at the distance $L$ to the earth.

\begin{figure}
\vspace{0mm}
\begin{center}
\includegraphics[width=8.5cm]{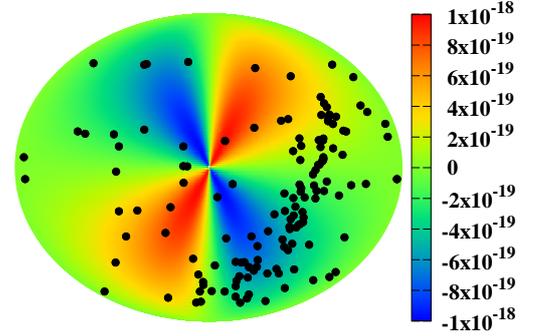}
\end{center}
\vspace{-8mm}
\caption{Spatial pattern of the bias factor $\alpha(\hat{\Omega},\hat{p},\theta)$ in the sky for $\dot{h}_+ = 10^{-18} s^{-1}$. The GW source position is placed at the center of the celestial sphere in equatorial coordinates.}
\label{alpha}
\end{figure}

In the following, we will discuss GWs with periods much longer than the observational time span, and in this case, the GW amplitude changes linearly with time. At the same time, we use the assumption that the second term ("pulsar term") is a random noise with zero average, which is reasonable when the GW wavelength is much shorter than the typical distance to pulsars. Thus, the GW frequency range we consider here is $10^{-12}{\rm Hz} \lesssim f_{\rm GW} \lesssim 10^{-10}$Hz. For such GWs, we can rewrite Eq.(\ref{e11}) as
\begin{eqnarray}
\Delta h_A (t^{\prime},\hat{\Omega},\theta) &=& \dot{h}_A(\hat{\Omega},\theta) t.
\label{e12}
\end{eqnarray}
Then, substituting Eq.(\ref{e11}) into Eq.(\ref{e1}), we find the timing residual induced by ultra-low frequency GWs is described by
\begin{eqnarray}
r_{GW}(t) &=& \frac{1}{2}\sum_{A = + , \times} F^A(\hat{\Omega}, \hat{p}) \dot{h}_A(\hat{\Omega},\theta)t^2.
\label{e13}
\end{eqnarray}
This time dependence is the same as timing residual induced by pulsar spin down, which is given by 
\begin{eqnarray}
r_{\dot{p}}(t) &=& \frac{1}{2}\frac{\dot{p}}{p}~t^2,
\label{e14}
\end{eqnarray}
where $p$ and $\dot{p}$ are the pulse period and spin-down rate, respectively. Therefore, the influence of ultra-low-frequency GWs is absorbed into the spin-down rate of the pulsar and cannot be identified in the standard analysis of PTA. On one hand, in the presence of ultra-low-frequency GWs, spin-down rate is biased as,
\begin{eqnarray}
\frac{\dot{p}_{\rm obs}}{p} &=& \frac{\dot{p}_0}{p} + \alpha(\hat{\Omega},\hat{p},\theta),
\label{e15}
\end{eqnarray}
where $\dot{p}_{\rm obs}$ and $\dot{p}_0$ are observed and intrinsic spin-down rates, respectively, and the bias factor $\alpha(\hat{\Omega},\hat{p})$ is given by,
\begin{eqnarray}
\alpha(\hat{\Omega},\hat{p},\theta) &=& \sum_{A={+},{\times}} F^{A}(\hat{\Omega},\hat{p}) \dot{h}_A(\hat{\Omega},\theta).
\label{e16}
\end{eqnarray}
The bias factor depends on the relative direction between the GW propagation and each pulsar and the spatial pattern is plotted in Fig.~\ref{alpha}.
Here, $\dot{h}_+(\Omega,\theta)$ and $\dot{h}_\times(\Omega,\theta)$ are depend on GW polarization $\theta$ and given by
\begin{eqnarray}
\dot{h}_+(\hat{\Omega},\theta) &=& \dot{h}(\hat{\Omega}) \cos{2\theta}, \\
\dot{h}_\times(\hat{\Omega},\theta) &=& \dot{h}(\hat{\Omega}) \sin{2\theta}.
\end{eqnarray}

In \citet{Yonemaru1}, we proposed a method to utilize the spatial pattern of $\alpha(\hat{\Omega},\hat{p},\theta)$ to probe ultra-low-frequency GWs. First, by assuming GW source position and polarization, we divide pulsars into two groups according to the sign of the bias factor, depending on the location of each pulsar.  Although GW signals cannot be extracted from individual pulsars, since spin-down rates are biased to positive and negative values in the two groups respectively, it is possible to detect GWs by measuring the systematic difference in the spin-down rate distribution between the two groups. 
We use the skewness of the spin-down rate distribution to characterize the bias induced by GWs, and convert the difference in the skewness of the two groups to the value of $\dot{h}$. Below, we apply this method to real data and derive constraints on $\dot{h}$ by analyzing the skewness difference. In addition, if the amplitude of GW is too strong, some of pulsars in the region with negative $\alpha(\hat{\Omega},\hat{p},\theta)$ get negative spin-down rate. We derive constraints on the GW amplitude by using the number of pulsars with negative spin-down rates in the real pulsar observation.

\section{Results}
\label{section3}

\subsection{Pulsar data}

We use data of observed MSPs in ATNF pulsar catalog ver. 1.59. The data set includes 181 MSPs with the measured periods shorter than 30 msec and the time derivatives. Among 181 MSPs, we exclude 30 MSPs in globular clusters since they would be biased significantly by the gravitational potential and complicated dynamics inside the cluster. In addition, two MSPs are removed as outliers: one with a negative spin-down rate $(\dot{p}_{\rm obs}/p = -10^{-20.2}~[{\rm sec}^{-1}]$,
J1801-3210) and one with an exceptionally large spin-down rate $(\dot{p}_{\rm obs}/p = 10^{-11.5}~[{\rm sec}^{-1}]$, J0537-6910). Thus, 149 MSPs are used for our analysis below.

\begin{figure}
\vspace{0mm}
\begin{center}
\hspace*{0mm}
\includegraphics[width=7cm]{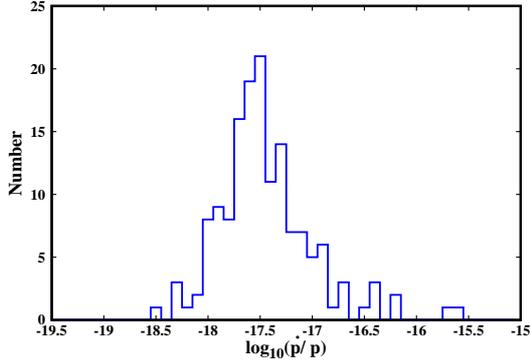}
\end{center}
\vspace{0mm}
\caption{Histogram of spin-down rates of 149 MSPs.}
\label{histoall}
\end{figure}

Fig.~\ref{histoall} shows the histogram of $\dot{p}_{\rm obs}/p$ of 149 MSPs. Mean, standard deviation, skewness and kurtosis of the distribution are $-17.4$, $0.36$, $1.2$ and $5.9$, respectively, and the deviation from Gaussian distribution was shown to be statistically significant by the Jarque-Bera test \citep{Yonemaru2}.

\subsection{GW search}
\label{section31}

\begin{figure}
\vspace{0mm}
\begin{center}
\hspace*{0mm}
\includegraphics[width=8cm]{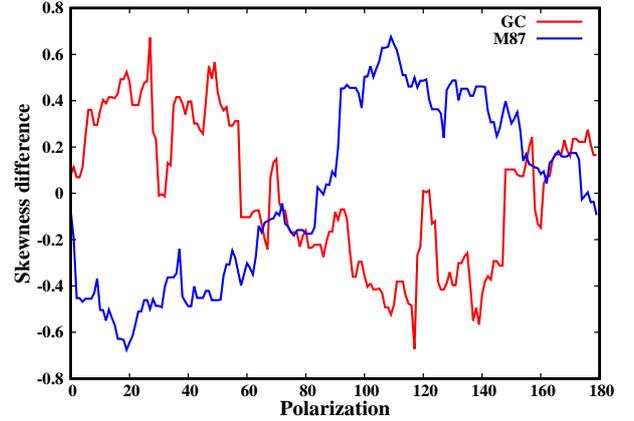}
\end{center}
\vspace{0mm}
\caption{Skewness difference as a function of GW polarization angle for the cases where the location of the GW source is assumed in the direction of the Galactic Center (red) and M87 (blue).}
\label{skewness_polarization}
\end{figure}

Given the propagation direction and polarization of GWs, the sky is divided into two areas according to the sign of the bias factor $\alpha(\hat{\Omega},\hat{p},\theta)$. Then MSPs are classified into two groups and skewness of the $\dot{p}_{\rm obs}/p$ distribution is calculated for each group. The skewness in the positive (negative) $\alpha(\hat{\Omega},\hat{p},\theta)$ region is given by
\begin{eqnarray}
S_{\alpha +(-)} &=& \frac{1}{\sigma^3_{+(-)}N_{+(-)}}\sum^{N_{+(-)}}_i \left( \log_{10}\left(\frac{\dot{p}_{\rm{obs}}}{p}\right)_i - \mu_{+(-)} \right)^3.
\end{eqnarray} 
where $i = 1,\cdots,N_{+(-)}$ is the number of MSP in the positive (negative) $\alpha(\hat{\Omega},\hat{p},\theta)$ region, and $\mu_{+(-)}$ and $\sigma^2_{+(-)}$ are the mean value and variance of the $\log_{10}\dot{p}/p$ distribution, respectively, 
\begin{eqnarray}
\mu_{+(-)} &=&  \frac{1}{N_{+(-)}}\sum^{N_{+(-)}}_i \log_{10}\left(\frac{\dot{p}_{\rm{obs}}}{p}\right)_i, \\
\sigma_{+(-)}^2 &=& \frac{1}{N_{+(-)}}\sum^{N_{+(-)}}_i \left( \log_{10}\left(\frac{\dot{p}_{\rm{obs}}}{p}\right)_i - \mu_{+(-)} \right)^2. 
\end{eqnarray}
Then the skewness difference is given by
\begin{eqnarray}
\Delta S &=& S_{\alpha +} - S_{\alpha -}.
\end{eqnarray}
Fig.~\ref{skewness_polarization} shows the skewness difference as a function of GW polarization angle for the cases where we assume that the GW source is located in the direction of the Galactic Center and M87. It should be noted that polarization angles of $0$ and $90$ degree correspond to the same polarization, but the sign of $\alpha$ is inverted so that the sign of skewness difference is also inverted. Maximum values for the case of the Galactic Center and M87 are 0.672 and 0.676, respectively.

\begin{figure}
\vspace{0mm}
\begin{center}
\hspace*{0mm}
\includegraphics[width=8.5cm]{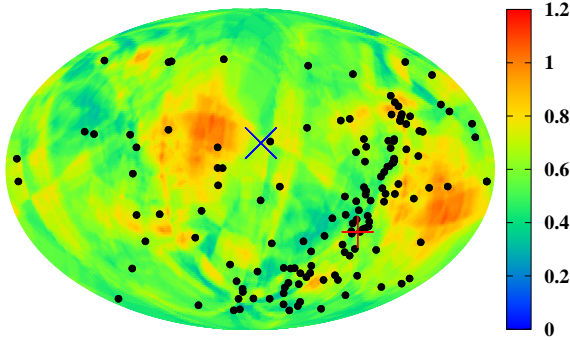}
\end{center}
\vspace{-8mm}
\caption{Skewness difference distribution in the sky. At each point of the sky, where the GW source is assumed to be, we compute the skewness difference by varying GW polarization angle and the colors represent the maximum skewness difference at each point. Black points represent the position of MSPs. Red "+" and blue "$\times$" represent the position of the Galactic Center and M87, respectively.}
\label{GWsearch}
\end{figure}

In the same way, we calculate the skewness difference for all directions of GW source and all angles of GW polarization. In Fig.~\ref{GWsearch}, at each point of the sky where GW source is assumed to be, we have searched for the largest skewness difference by changing the GW polarization. The skewness difference is mostly smaller than unity and the largest value in the whole sky is $1.07$. The two red regions are antipodes.

In order to study the statistical significance of this value, we perform a series of simulations. First, we make mock data of 149 MSPs in absence of GWs. MSPs are located at the same position as that of the real data. The value of $\dot{p}_{\rm obs}/p$ is randomly allocated to each MSP according to a generalized Gaussian distribution with the same values of mean, standard deviation and skewness as the ones obtained from the catalog. The generalized Gaussian distribution is given by 
\begin{eqnarray}
f(x) &=& \frac{\phi(y)}{\alpha - \kappa(x - \xi)},
\label{e17}
\end{eqnarray}
where $\phi(y)$ is the standard Gaussian distribution and $y$ is given by
\begin{eqnarray}
y &=& \left\{ \begin{array}{ll}
-\frac{1}{\kappa} \log[1 - \frac{\kappa(x - \xi)}{\alpha}] & (\kappa \neq 0) \\
\frac{\kappa(x - \xi)}{\alpha} & (\kappa = 0) ,
\end{array} \right.
\label{e18}
\end{eqnarray}
Here, $\xi$, $\alpha$ and $\kappa$ are the location, scale and shape parameters, respectively, and the mean $\mu$, standard deviation $\sigma$ and skewness $S$ are expressed by these parameters.
\begin{eqnarray}
\mu &=& \xi - \frac{\alpha}{\kappa} \left(\mathrm{e}^{\kappa^2 / 2} - 1\right),\\
\label{e19}
\sigma &=& \sqrt{\frac{\alpha^2}{\kappa^2}\left(\mathrm{e}^{\kappa^2} - 1\right)},\\
\label{e20}
S &=& \frac{3 e^{\kappa^2} - e^{3 \kappa^2} - 2}{(e^{\kappa^2} - 1)^{3/2}} {\rm sgn}(\kappa).
\label{e21}
\end{eqnarray}
For each realization of mock data, we obtain the skewness difference in the same way as above and search for the maximum varying the position and polarization.

\begin{figure}
\vspace{0mm}
\begin{center}
\hspace*{0mm}
\includegraphics[width=8cm]{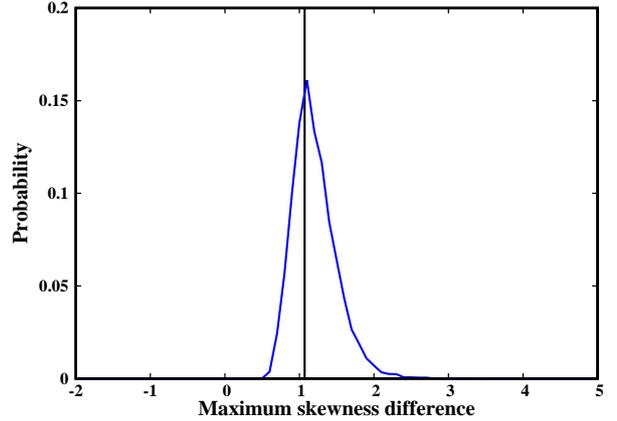}
\end{center}
\vspace{0mm}
\caption{Simulated probability distribution function of the maximum skewness difference in the sky, which is
obtained through 10,000 realizations of mock MSP data. The vertical line shows the value $1.07$ obtained from the real MSP data.}
\label{significant}
\end{figure}

Fig.~\ref{significant} shows the probability distribution function of the maximum skewness difference in the sky obtained through 10,000 realizations of mock MSP data without GW injection. We find that the distribution extends from $0.6$ to $2.0$ and, as a result, the obtained value $1.07$ is fairly consistent with the statistical fluctuations without GWs.

\subsection{GW limits from skewness difference}
\label{section32}

In the previous subsection, we have shown that the current pulsar data is consistent with the non-existence of GWs within the statistical error. In this subsection, we derive upper bounds on the derivative of the GW amplitude $\dot{h}$. Due to the limited computational power, we focus on two astrophysically important directions of Galactic Center and M87 where the existence of supermassive black hole binaries has been suggested \citep{Yu,Oka,Yonemaru1}.

As we saw in the previous subsection, the maximum skewness differences in the direction of the Galactic Center and M87 are 0.672 (at polarization angle of 25 deg) and 0.676 (at polarization angle of 108 deg), respectively. In the presence of GWs with large enough value of $\dot{h}$, the probability of obtaining such small values is low. We place the upper bound on $\dot{h}$ by using the threshold where the probability of having skewness difference less than 0.672 (0.676) is $2\%$ for the direction of the Galactic Center (M87).

In order to evaluate the upper bounds, we make mock data of 149 MSPs in the same way as the previous subsection and inject the bias due to GWs to the simulated intrinsic spin-down rates $\dot{p}_0/p$. The polarization angle of GWs is set to be the same as the above: 25 deg for the Galactic Center and 108 deg for M87. In this way, a simulated value of $\dot{p}_{\rm obs}/p$ is allocated to each MSP and then we compute the skewness difference for the data set. It should be noted that MSPs with very small values of intrinsic spin-down rate ($\sim 10^{-18}~{\rm sec}$) can have negative values of observed spin-down rate if they are located at an area with negative bias. They are removed from the computation of skewness difference and the total number of used MSPs is slightly smaller than 149.

\begin{figure}
\vspace{0mm}
\begin{center}
\hspace{0mm}
\includegraphics[width=8cm]{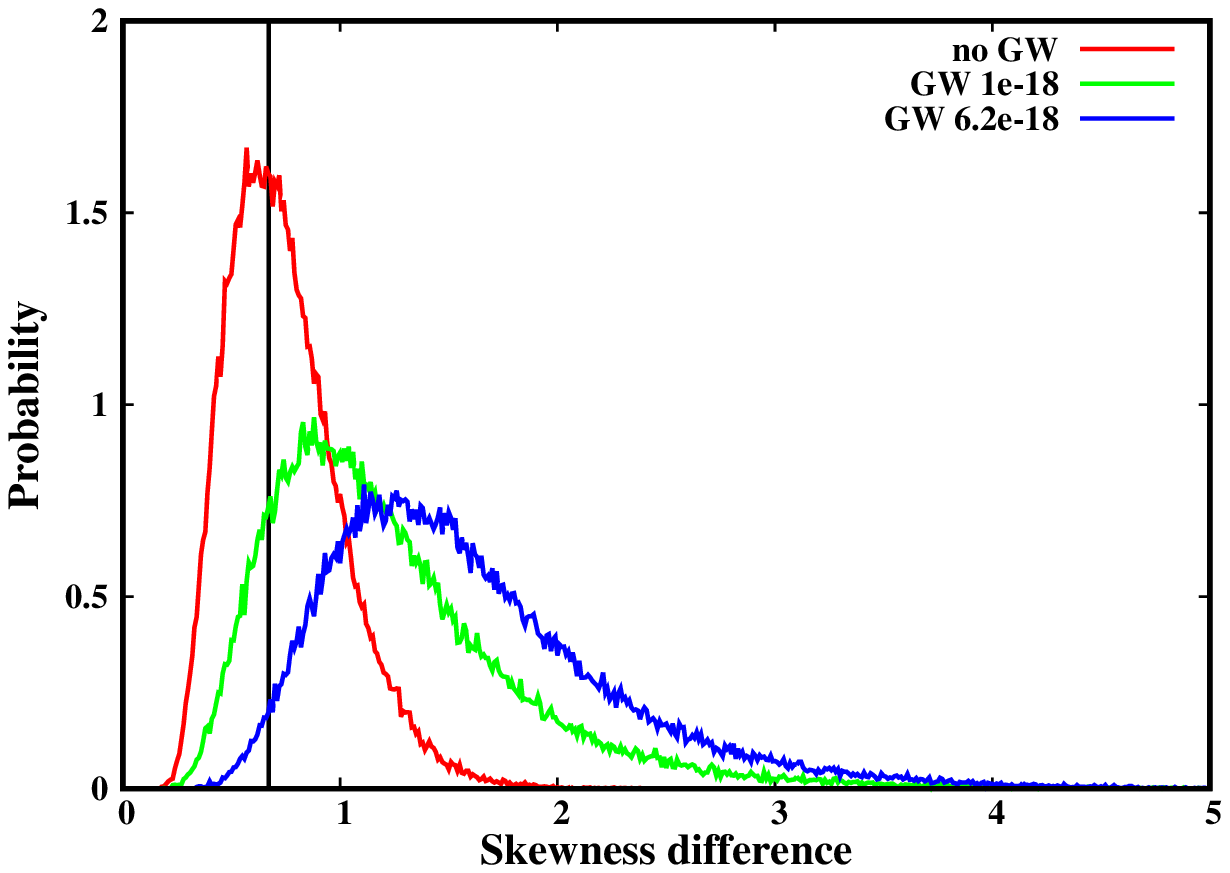}
\includegraphics[width=8cm]{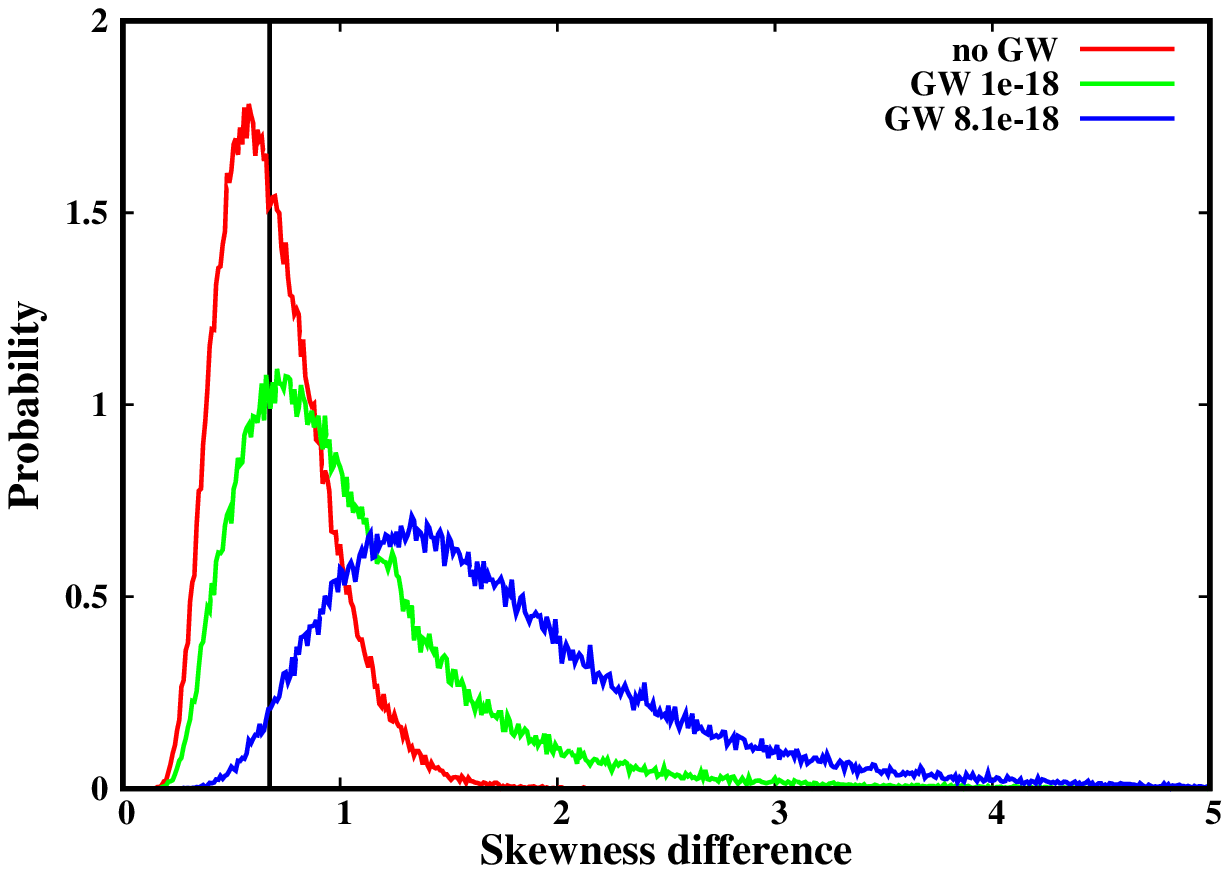}
\end{center}
\vspace{0mm}
\caption{Simulated probability distribution function of skewness difference in the directions of the Galactic Center (top) and M87 (bottom) estimated from 10,000 realizations of simulations without GWs (red) and with GWs (green and blue). The value of $\dot{h}$ is $10^{-18}~{\rm sec}^{-1}$ (green) and $6.2 \times 10^{-18}~{\rm sec}^{-1}$ (blue) for the Galactic Center, and $10^{-18}~{\rm sec}^{-1}$ (green) and $8.1 \times 10^{-18}~{\rm sec}^{-1}$ (blue) for M87.}
\label{histoGW}
\end{figure}

Fig.~\ref{histoGW} shows the probability distribution function of skewness difference in the directions of the Galactic Center and M87 estimated from 10,000 realizations of simulations with and without GWs. Here, we fix GW polarization angles at $25~{\rm deg}$ (GC) and $108~{\rm deg}$ (M87), which give the largest skewness difference in Fig.~\ref{skewness_polarization}. We see that, as the value of $\dot{h}$ increases, the probability of having a larger value of skewness difference becomes higher. Comparing them with the observed values (0.672 and 0.676), we obtain an upper bound of $\dot{h} < 6.2 \times 10^{-18}~{\rm sec}^{-1}$ for the Galactic Center and $\dot{h} < 8.1 \times 10^{-18}~{\rm sec}^{-1}$ for M87. The implication of the upper bounds will be discussed in Section.~\ref{section4}.

\subsection{GW limits from the number of MSPs with negative spin-down rates}
\label{section33}

As we mentioned in the previous subsection, when MSPs with very small intrinsic spin-down rates are biased negatively, they can have negative values of observed spin-down rate. The number of such MSPs will increase for stronger GWs. Therefore, the number of MSPs with negative observed spin-down rates could be used for another measure to probe ultra-low-frequency GWs. In the current data set, there is only one MSP with a negative $\dot{p}_{\rm obs}/p$ except ones in globular clusters. Here, we set upper bounds on $\dot{h}$ using the threshold where the probability of having two or more MSPs have negative $\dot{p}_{\rm obs}/p$ is $98\%$.

\begin{figure}
\vspace{0mm}
\begin{center}
\hspace{0mm}
\includegraphics[width=8cm]{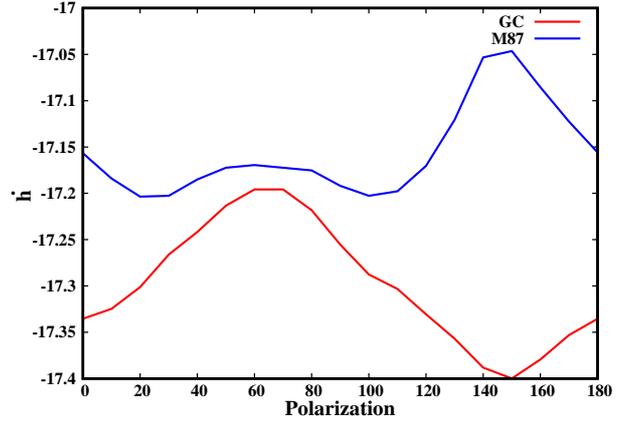}
\end{center}
\vspace{0mm}
\caption{Upper bounds on $\dot{h}$ from the number of MSPs with $\dot{p}_{\rm obs}/p$ as a function of GW polarization angle for the Galactic Center (red) and M87 (green).}
\label{imasato}
\end{figure}

Fig.~\ref{imasato} shows upper bounds on $\dot{h}$ as a function of GW polarization angle for the Galactic Center and M87. The upper bounds are of order $10^{-17}-10^{-17.4}~{\rm sec}^{-1}$ and comparable to those from skewness difference obtained in the previous subsection. GWs from the Galactic Center are slightly well constrained than those from M87 and the dependence on the polarization angle is very weak.

\section{Discussion}
\label{section4}

In the previous section, we have obtained upper bounds on the time derivative of GW amplitudes, rather than the amplitudes themselves. Using an approximation $\dot{h} \sim 2 \pi f_{\rm GW} h$ where $f_{\rm GW}$ is the frequency of GW, which is reasonable for most of the periods, our constraints for the Galactic Center and M87 can be approximated as,
\begin{eqnarray}
h_{\rm GC} &\lesssim& 3 \times 10^{-8} \left( \frac{1/1000~{\rm year}}{f_{\rm GW}} \right), \label{eq:bound-GC} \\
h_{\rm M87} &\lesssim& 4 \times 10^{-8} \left( \frac{1/1000~{\rm year}}{f_{\rm GW}} \right), \label{eq:bound-M87}
\end{eqnarray}
respectively.
On the other hand, the recent PTA analyses put upper bounds of $h_{GW} \sim 7.3 \times 10^{-15}$ at the frequency of 8 nHz and the bound scales as $f^{-2}_{\rm GW}$ \citep*{Aggarwal, Babak}. Thus, our constraints are comparable to those of standard PTAs at $f_{\rm GW} \sim 1 / 30000~{\rm years} \sim 10^{-12}~{\rm Hz}$ and are better at even lower frequencies. Although our constraints are weaker for $f_{\rm GW} \gtrsim 10^{-12}~{\rm Hz}$, they are still valuable as independent constraints. It should be noted that at frequencies lower than $10^{-13}~{\rm Hz}$, the pulsar term cannot be treated as random noise and pulsar distances are necessary to account for it \citep{Yonemaru2,Hisano}. Currently, the distance is not available for most pulsars and we do not consider this frequency range here. Thus, the constraints Eqs. (\ref{eq:bound-GC}) and (\ref{eq:bound-M87}) are applicable for $f_{\rm GW} \gtrsim 10^{-13}~{\rm Hz}$.

Let us consider possible SMBH binaries at these cites. A SMBH with mass of $4.0 \times 10^6M_{\odot}$ is known to reside in the Galactic Center and the possibility of the existence of another SMBH has been discussed (e.g. \citet{Oka}). If there exists a SMBH orbiting around the known SMBH, it could be a source of GWs. Assuming the period of the binary motion to be $100~{\rm years}$, the upper bound of Eq.(\ref{eq:bound-GC}) translates into a upper bound on the companion mass of $2 \times 10^{16}~M_{\odot}$.

Concerning the M87, the mass of a SMBH is estimated to be $6.6 \times 10^9 M_{\odot}$ at the center of M87. it has been indicated to have secondary SMBH and would be binary and GWs from such a potential pc-scale SMBH binary has been discussed  \citep*{Betcheldor,Yonemaru1}, while constraints on the amplitude of GWs emitted by a milli-pc scale SMBH binary in the PTA frequency bands has been already studied \citep{Schutz}. Assuming the orbital period to be $100~{\rm years}$, the upper bound of Eq.(\ref{eq:bound-M87}) results in a upper bound on the companion mass of $4 \times 10^{16} M_{\odot}$. 

In the above considerations, binaries are assumed to have circular orbits and inclination is zero degree (face-on).  In our previous works \citep*{Yonemaru2,Hisano}, we estimated future constraints on ultra-low-frequency GWs with 3,000 MSPs, which are expected to be found by the SKA2.
There, we found that GWs with $\dot{h}$ as small as $\sim 3 \times 10^{-19}~{\rm sec}^{-1}$ can be detected and the second SMBH mass as small as $3 \times 10^{14}~M_{\odot}$ could be probed in the case of circular orbits and zero inclination. In fact, the GW amplitude is sensitive to the eccentricity and the phase of the binary motion.


\begin{figure}[t]
\vspace{0mm}
\begin{center}
\hspace{0mm}
\includegraphics[width=9cm]{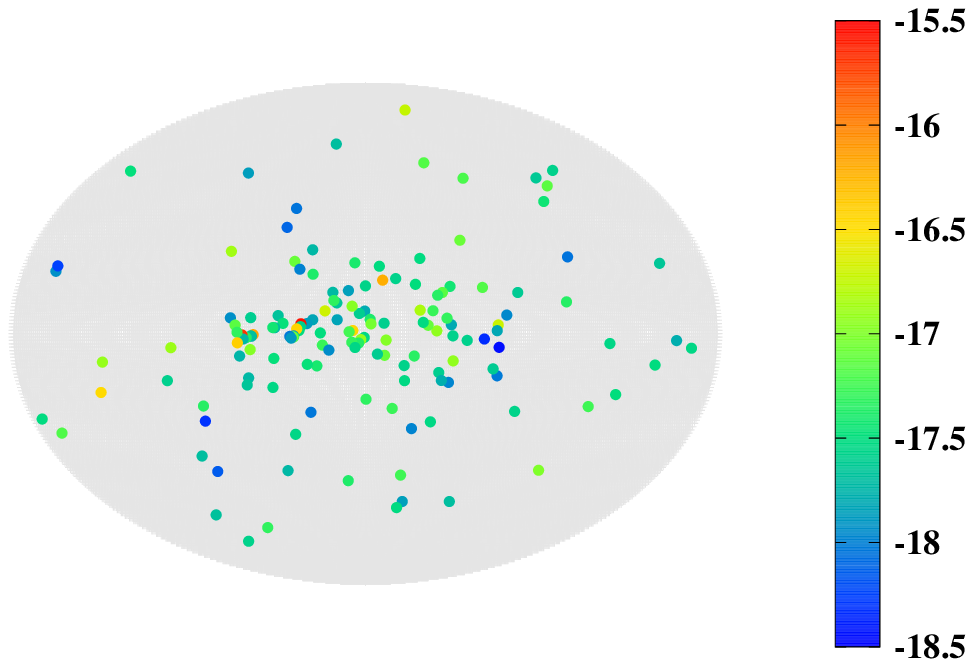}
\includegraphics[width=8cm]{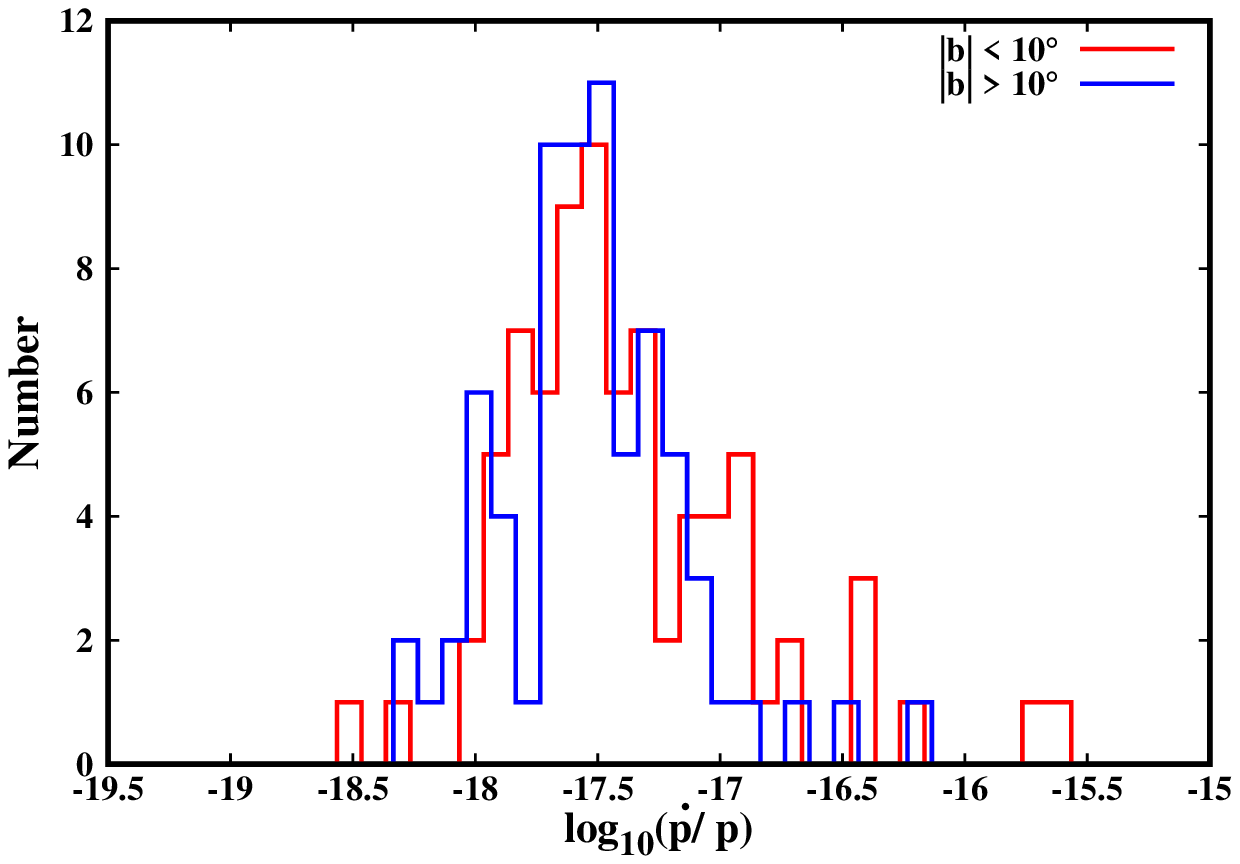}
\end{center}
\vspace{0mm}
\caption{(Top) Position of 149 MSPs in the galactic coordinates with the spin-down rates represented by the color. (Bottom) Histogram of spin-down rates for MSPs in the Galactic plane ($|b| < 10~{\rm deg}$, red) and outside ($|b| > 10~{\rm deg}$, blue).}
\label{galactic}
\end{figure}

Our simulation is based on the assumption that the distribution of spin-down rate in the sky is isotropic in the absense of GWs. However, considering the evolution of MSPs, it may not be the case. Neutron stars are mostly produced in Galactic plane and often have large peculiar velocities due to the kick at supernovae \citep{Hansen}. Since MSPs with small values of $\dot{p}/p$ have long characteristic age ($p/\dot{p}$), they may tend to be located far from the birth place outside the Galactic plane. Thus, there is a possibility that MSPs with large (small) $\dot{p}/p$ are populated outside (inside) of the Galactic plane, which induces the anisotropy of $\dot{p}/p$ distribution in the sky and results in systematics in our method.

In the top panel of Fig.~\ref{galactic}, the position of 149 MSPs in the galactic coordinates is shown with the indication of spin-down rates. In the bottom panel, we show the histogram of spin-down rates of MSPs within Galactic plane ($|b| < 10~{\rm deg}$) and outside ($|b| > 10~{\rm deg}$), separately. The mean and standard deviation of the histograms are -17.4 and 0.5 for $|b| < 10~{\rm deg}$, and -17.5 and 0.4 for $|b| > 10~{\rm deg}$, respectively. Thus, significant difference is not found between the two histograms and the systematics is considered to be negligible. Here it should be noted that MSPs with $|b| > 10~{\rm deg}$ may reside within the Galactic disk but it cannot be known without information on the distance.

Finally, we note that the observed spin-down rates could be biased by other factors than GWs such as the Shklovskii effect \citep{Shklovskii}, the Galactic differential rotation \citep*{Damour, Rong} and acceleration toward the Galactic disk \citep{Nice}. Although the biases from the Galactic differential rotation and acceleration toward the disk would have spatial correlations in the sky, these effects are less significant ($\Delta(\dot{p}/p)$ $\leq 10^{-19}$ for MSPs at $\leq $ 10 kpc) \citep{Nice} and will be removed if the distance to MSPs is measured precisely in the future.

\section{Summary}
\label{section5}

In this paper, we have placed constraints on GWs from a single source with ultra-low frequencies ($10^{-12}~{\rm Hz} \lesssim f_{\rm GW} \lesssim 10^{-10}~{\rm Hz}$) by applying a method proposed in \citet{Yonemaru1} to observed milli-second pulsars (MSPs). This method is based on the statistics of spin-down rate distribution, where the skewness difference between two MSP groups divided according to the position in the sky, is used to find a bias induced by GWs. We selected 149 MSPs from the ATNF pulsar catalog and calculated the skewness difference. By comparing with mock MSP data, we have shown that the current MSP data is consistent with no GWs from any direction of the sky. Furthermore, we have derived upper bounds on the time derivative of the GW amplitude $\dot{h} < 6.2 \times 10^{-18}~{\rm sec}^{-1}$ and $\dot{h} < 8.1 \times 10^{-18}~{\rm sec}^{-1}$ for the Galactic Center and M87, respectively. Consistent bounds were derived from the number of MSPs with negative spin-down rates. Approximating the GW amplitude as $\dot{h} \sim 2 \pi f_{\rm GW} h$, the bounds respectively translate into $h < 3 \times 10^{-9}$ and $h < 4 \times 10^{-9}$ for $f_{\rm GW} = 1/(100~{\rm yr})$. The constraints will improve by more than one order of magnitude with 3,000 MSPs in the SKA era \citep{Yonemaru2,Hisano}.

\section*{Acknowledgements}
We thank George Hobbs for useful discussion. NY was financially supported by the Grant-in-Aid from the Overseas Challenge Program for Young Researchers of JSPS. SK is partially supported by the Grant-in-Aid for Scientific Research from JSPS, Grant Number 17K14282, and by the Career Development Project for Researchers of Allied Universities. KT is partially supported by JSPS KAKENHI Grant Numbers JP15H05896, JP16H05999 and JP17H01110, and Bilateral Joint Research Projects of JSPS.
The Parkes radio telescope is part of the Australia Telescope which is funded by the Commonwealth of Australia for operation as a National Facility managed by CSIRO.
The ATNF pulsar catalogue at \href{http://www.atnf.csiro.au/people/pulsar/psrcat/}{http://www.atnf.csiro.au/people/pulsar/psrcat/}
was used for this work.
We also thank the referee for a careful reading of the manuscript.


\bsp	
\label{lastpage}
\end{document}